\documentclass[12pt]{article}
  \usepackage{a4}
  \newcommand\ignore[1]{}
  \usepackage{epsfig,graphicx}

\begin{document}
 \hfill To be published in Proceedings of ECRYS-99,

\hfill Journ. de Physique, Coll., December 1999.

\vskip .6in

  {\Large{\bf{ Theory of plastic flows of CDWs in application to a current conversion.}}}
 
 \vskip .1in
 {\centerline{S. Brazovski$^{1,2}$,  N. Kirova$^{1}$.}
  \vskip .1in

{\it{ $^1$Laboratoire de Physique Th\'eorique et des Mod\`eles Statistiques, CNRS,}

\it{$^{\ }$B\^at.100, Universit\'e Paris-Sud, 91406 Orsay cedex, France.}

\it{$^2$L.D. Landau Institute,  Moscow, Russia. }}

  \vskip .2in
 
  \begin{narrow}
 
  {\bf Abstract.}
   We suggest  a theoretical picture for distributions of plastic deformations experienced  by a sliding Charge Density Wave  in the course of the   conversion from the normal current at the contact to the collective  one in the bulk. Several mechanisms of phase slips via  creation and proliferation of dislocations are compared. The results are applied to space resolved X-ray [1,2],  multi-contact [3,4] and optical [5] studies. Numerical
 simulations are combined with model independent relations.
  \end{narrow}
 
\vskip .2in

{\bf 1. INTRODUCTION.}\\

\vskip .1in
 Plastic deformations and flows are currently the focus of studies of
Electronic Crystals (charge/spin density waves(DW) and
interface 2d Wigner crystals) and of vortex lattices.
Thus, dislocations are supposed to participate in depinning, in phase
slips, in narrow band noise generation, and also arise in contact structures.
Experimental studies show that the sliding state of DWs is essentially  inhomogeneous [1-6]  with different effects near contacts and in the bulk. We  present
 general  scheme
to describe
 inhomogeneous distributions of 
deformations (which are the gradients $\varphi^{\prime}=q $ 
along the chain axe $x$ of the phase $\varphi$), electric fields and currents in the course of the  current conversion.   For a long enough sample we predict a strong variation for the shift  $q(x)$
of the CDW wave number near contacts which flattens (nearly exponentially) towards the bulk, 
$\delta x\to\infty$.The bulk strain like  $q=q_{blk}^{\prime}x$ remains only when the conversion is not able to bring the partial currents to equilibrium  with strains allowed at this expense. It happens either for  size dominated configurations like in [3], or when the pinning of dislocations (DLs) hinders the conversion process as demonstrated by X-ray experiments [1]. We present numerical modeling for time dependent and stationary cases.
\vskip .2in
{\bf 2. MODEL INDEPENDENT RELATIONS.}
\vskip .1in
 Generally in the CDW state there are two types $\alpha = i,e$ of normal carriers which share  the metallic state  DOS $N_F$ ( per chain area $s$).
For intrinsic ones, $\alpha = i$,  the partition 
$N_F^i=\beta_i N_F$  is closed below $T_p$, so that they are only excited across the Peierls gap.
The extrinsic carriers are present in semimetalic DWs: $NbSe_3$ and some SDWs.
Their  DOS partition $\beta_e N_F$ remains open, $\beta_i +\beta_e =1$.
The potentials $V_\alpha$ experienced by the carriers and their  chemical potentials $\mu_\alpha$ are [7,8]
\[
V_e = \Phi \ , \ V_i = \Phi + q/(\pi N^i_F) \, ; \,
\mu_\alpha \approx V_\alpha + \delta n_\alpha/(\rho_\alpha N^\alpha_F)
\]
where $\Phi$ is the electric potential and $\delta n_{\alpha}$ are the local imbalances between electron and hole concentrations for each type of carriers.
The normalized normal and condensate densities, $\rho_i(T)$ and $\rho_c(T)=1-\rho_i(T)$, change from $\rho_i = 1$, $\rho_c = 0$ in the metallic state to $\rho_i = 0$, $\rho_c = 1$ at low $T$ while $\rho_e\approx 1$ stays constant.
Fast equilibration of quasi-particles leads  to 
$\mu_i = \mu_e \equiv \mu_n$ and the normal current is $j_n = -(\sigma_n s/e^2) \partial\mu_n/\partial x$ where 
$\sigma_n = \sigma_i + \sigma_e$ is the normal conductivity.
(All quantities are given per electron per chain, hence the conductivities acquire the factor $s/e^2$.)

We separate the total  densities of the charge $\delta n_{\rm tot}$ and of  the current $j_{\rm tot}$ into their normal and condensate counterparts, 
$\delta n_{\rm tot} = \delta n_n + \delta n_c$ and $j_{\rm tot} = j_n + j_c$, with $\delta n_n = \delta n_e + \delta n_i$, $\delta n_c = \varphi^\prime/\pi$ and $j_c = -\dot\varphi /\pi$. (The renormalization $\delta n_c\rightarrow \rho_c \varphi^\prime/\pi$ appears inherently [8] as an action of the potential $V_i$.)
The stress $2U$, per DW period, includes [7] three contributions: elastic, electronic and electric. \begin{equation}
U = (q/\pi + \delta n_i)/N^i_F + \Phi \ ,\ 
\partial U/\partial x \equiv -{\cal F} ={\cal F_{\rm frc}}
\label{CDWstr}
\end{equation}
The electronic term $\sim \delta n_i$ in $U$ is conjugated to the term $\sim q$ in $V_i$. Both come from corresponding variations of the breathing energy $\delta n_i q/(\pi N_F^i)$. As a variation of the energy over $\delta n_c$, the potential  $U$ plays a role of the condensate  potential $V_c$. 

The second equation in (\ref{CDWstr}) states that the driving force of the DW sliding  ${\cal F}$ is equilibrated by the friction force 
$-{\cal F}_{\rm frc}(j_c)$ which coincides with electric field ${\cal E}$ in equilibrium ($q=0,\ \delta n_i=0$). 
Thus the function  ${\cal F}(j_c)$  can be obtained from the sliding $V-I$  characteristics for distant nonperturbative contacts. We always suppose that ${\cal F}$ and $j_{tot}$ are above the sliding threshold values ${\cal F}_t$ and $j_{t}=(\sigma_n s/e^2){\cal F}_t$.
The local electro-neutrality condition [9] implies that $\delta n_n$ images $q$  or vice versa:  $-\delta n_n = \delta n_c \equiv  q/\pi$ thus linking X-ray [1,2]  and optical [5] experiments.
With the above relations  we can express $q$ and $\cal E$ as:

\begin{equation}
qg/\pi N^i_F = \mu_n-U\equiv\eta \qquad {\cal E}={\cal F}(j_{c})-q^{\prime}\beta _{e}g/\pi N^i_F
  ; \ 
g^{-1}= \beta_e +\rho_i/\rho_c
\label{EtaEquation}
\end{equation}
Here $g=g(T)$ is the normalized elastic modulus with $\beta_e$ and $\rho_i$ characterizing the Coulomb screening of the CDW deformations by  carriers. 
(In other theories only $g=\rho_c $ has been exploited which is valid  only near  $T_c$.)

We see that  only via extrinsic carriers the elastic stress enters the balance between the forces. Otherwise $\cal E$ follows ${\cal F}(j_c)$ identically.
The first relation in (\ref{EtaEquation}) shows that the CDW is deformed whenever the stress $U$  does not coincide with the chemical potential of the normal carriers $\mu_n$. The quantity $\eta$  measures this mismatch, and hence characterizes an excess or a lack of normal carriers which is imaged directly by $q$.
The exchange between the normal carriers and the condensate via the phase slips finally equilibrates $\mu_n$ and $U$, a process taking place via nucleation and growth of DLs. The asymptotic condition for the bulk is that  {\em all partial currents are stationary} which implies the   absence of current conversion, all types of carriers being in equilibrium, with  equal chemical potentials hence $\eta\to 0$.   Only {\em {hindering of phase slips}} can prevent leveling of $\eta$, and hence $q$, to zero far from a contact.

The population of normal carriers is controlled by the injection/extraction rates 
$\nu _{inj}=\pm j_{tot}\delta (x-x_{cnt})$ from the contacts and by the conversion rate $2{\cal R}$ to/from the condensate:
$dn_n/dt\equiv \dot n_n+j^\prime_n=\nu _{inj}-2{\cal R}$. 
Here  
${\cal R}={\cal R}(\eta,j_c)$ is the rate of phase slips  (per unit length per chain) which transform pairs of electrons to/from the DW periods. (See [10,11] for a review of microscopic mechanisms.)  With the above relations  we arrive at the two equations for two variables  $\eta$ and $j_c$:
\begin{equation}
{\dot\eta \pi N_F^i}/g+j^\prime_c-2{\cal R}(\eta,j_c )+\nu_{\rm inj}=0 \qquad
{\cal F}(j_{c})-(e^2/s\sigma _{n})(j_{tot} -j_c) 
=\eta^{\prime } 
\label{eta-U}
\end{equation}
They
 provide a general scheme for the plastic sliding with time and space dependent current conversion processes. There are common elements with the modeling scheme used in [3]. The major difference is that our equations provide the flat asymptotics $q\rightarrow 0$ and also Eq. (2) suggests a different relation between the measured $\cal E$ and the extracted $j_c$. Next, all dynamics is generated by a nonlinear function ${\cal{F}}(j_c)$ while in [3] its linear part (dominating at $j_c \gg j_t$ only) was extracted, the rest being treated as an a posteriori correction. Further on, our form of $g$ allows to study the semiconducting regime.

The most sensitive element of any construction is the function ${\cal R}(\eta,j_c )$ which may differ in various circumstances. This function  is of great interest itself so that the above Eqs. can be utilized to recover it from measurements of $q$ or ${\cal E}$. Otherwise we can guess its form in plausible pictures of the current conversion to get $q,\ {\cal E}$ and to compare those with experiment. 
The $HomoN$ scenario refers to an ideal host crystal, both in the bulk and at the surface, where only homogeneous nucleation  is present as a spontaneous thermal [12] or even quantum [13] supercritical fluctuation, so that ${\cal R}_{\rm hom}\propto\exp[-\eta_0/|\eta|]$, valid at $|\eta| \ll \eta_0$. In general the $HomoN$ scenario is very unlikely in any physical system as requiring an extreme purity; also in our problem it cannot provide an observed fast contact variation of $q$.

The $HeteroN$ scenario refers to samples with a sufficiently large density of defects acting as nucleation centers for supercritical DLs, the simplest form for this case being ${\cal R}_{\rm het} \approx R_p N_{F}^{i}\eta$. Since $N_{F}^{i}\eta/g$ is an over-saturation of the electronic concentration then 
$gR_p=\tau_{cnv}^{-1}$ is nothing but the inverse life time of an excess  normal carrier with respect to its absorption by a DL.
 In DWs the heterogeneous nucleation will be provided by any region where $q$ varies over the cross-section which must be encircled by a sequence of DLs, one per a length of $2\pi$ increment. 

Both scenarios are of the $passive$ type, when the DW motion itself plays no role. 
A plausible $active$ scenario emerges for a fast enough DW motion when the DLs are created by the DW sliding through bulk or surface defects:
${\cal R} = {\cal R}_a(\eta,j_c)$ such that 
${\cal R}_a(0,j_c) = {\cal R}_a(\eta,0) = 0$ with ${\cal R}_a \approx R_a N_F^i\eta j_c$ as the simplest version. The origination of DLs at the crystallographic steps interrupting the DW sliding in a fraction of the sample cross-section is a most apparent manifestation. The origination of DLs pairs at defects in the bulk have been described in [10,14]  within the theory of the nonlinear  
${\cal F}_{frc} (j_c)$. An extension of this theory to describe also $\cal R$ would bring the whole complex of sliding effects to one frame.

At lowest $\vert \eta \vert$ the conversion bottleneck is a pinning of DLs by defects similar to vortex lines. It results either in a complete freezing of their motion or in a drastic reduction of their climb speed. Introducing $\eta_t$ as such a threshold or a crossover we have the following guesses:
i.collective pinning of DLs: ${\cal R}(\eta)=0$ at $\vert\eta\vert <\eta_t$;
ii. local pinning of DLs: ${\cal R}(\eta)=R_pN_F^i \eta \exp (-\eta_t/\vert \eta \vert)$
The last law for the DLs climb provides a logarithmic time decay  of $q$ observed  near contacts in pulsed experiments of [1], similar to the Kim-Anderson law in superconductors.
\vskip .2in
{\bf 3. APPROXIMATE SOLUTIONS AND MODELING.} \\
\vskip .1in
For an illustration consider a linear  sliding  $I-V$:
$\sigma_{c}^{-1}=\partial({\cal F}/\partial j_c={\rm cnst}$. 
(Actually it is true only asymptotically, for currents much above $2J_t$ as in  experiments of [1] and even above $5J_t$ as in [3].)
Then for the $HeteroN$ model we arrive at the single linear Eq.:

\begin{equation}
\dot \eta/g-\eta^{\prime\prime}r_D^2\sigma^* +
2\eta {\cal R}_p=\pm\delta (x-x_{cnt})j_{tot}/ N_F^i \qquad r_D^{-2}=4\pi e^2N_F^i/s
\label{eta-t}
\end{equation}
where $\sigma^*=\sigma_n\sigma_c/(\sigma_n+\sigma_c)$ and $r_D\sim 1\AA$ is the screening radius (by intrinsic carriers) in the metallic phase.
This equation defines the characteristic time $\tau= (2 g{\cal R}_p)^{-1}$ and length $\lambda=r_D( 2\pi{\sigma^*}/{\cal R}_p)^{1/2}$ scales.
For the  local "phase slip voltage" $V_{ps}=\Phi(x,t)-xj_{tot}/\sigma$ we find
\begin{equation}
V_{ps}(x,t)=
q(x,t)\frac{g\sigma^*}{\pi N_F^i}\left(\frac{\beta_e}{\sigma_n}-\frac{\beta_i}{\sigma_c}\right)
\label{Vps-linear}
\end{equation}
The Eq.(\ref{Vps-linear}) shows that  $V_{ps}$ indeed follows the strain $q$ but with an unexpected coefficient, which may change the sign as a function of $T$ or $j_{tot}$ at 
$\beta_e=\sigma_n/(\sigma_n +\sigma_c)$. Notice that $i-$ and $e-$ carriers contribute oppositely to the contact electric field, see e.g. [8].

The apparent analytical solution of  Eq. (\ref{eta-t}) allows for easy
 time dependent simulations e.g. for alternating pulses of [3]. Fig.1 shows the evolution of $q(x)$ with time after the change of  the current polarity. We conclude that the non-monotonic behavior found in [3] is universal with respect to the conversion mechanism.

\begin{figure}[thb]
\includegraphics{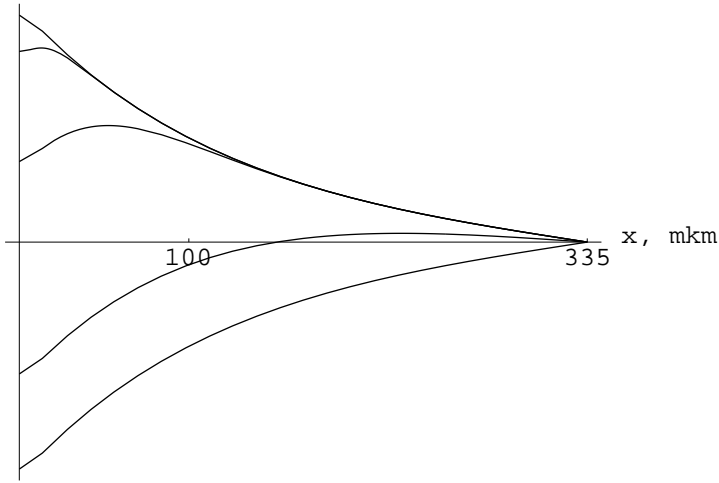}
\epsfxsize \hsize
\caption{Time dependent solutions (parameters from  [3]).}
\end{figure}

For  stationary distributions the same linear model gives for 
$x_{cnt}=\pm a$
\begin{equation}
q=-\frac{1}{4g}\frac{j_{tot}}{\sigma_n}\frac{\lambda}{r_D^2}\frac{\sinh (x/\lambda)}{\cosh (a/\lambda)}.
\label{q-linear}
\end{equation}
Its fitting, corrected for a small cutoff $\eta_t$, has been proved [1] to be very good.
The Eq.(\ref{q-linear}) shows that for a long enough sample there will be no linear gradient in the bulk.
The carrier conversion will be complete and the CDW deformation reduced to zero.
Otherwise for close, $a<\lambda$, contacts of experiments [3]\ we find a linear variation $q^\prime=j_{tot}/(4g\sigma_nr_D^2) $ corresponding to the  essentially non-equilibrium regime with a negligible  current conversion.
For the nonlinear $homoN$ model also from the analytical solution  we find near one contact: $q\propto (\ln|x-a|)^{-1}$.
This is an extremely slow decrease which results in an {\em incomplete carrier conversion} and in a large  strain along any realistic sample length, which contradicts to later  X-ray [1] and multicontact [4,5] experiments. 
 A reasonable fitting with this form of  ${\cal R}$  [2,3,12]  was due to its utilization well beyond the validity range 
$\vert \eta \vert \ll \eta_0$ when actually one should change to power laws.

The Fig.2 shows the distribution $2q(x)$ for $j=2.13j_t$ plotted against the
double shift data of dc experiments [1]. The $x$--axis gives the
distance from the left contact at $x=-2mm$ to the mid point  $x=0$.  Taking from the  data
the contact value of $2q_{\rm cnt}=12.6\times10^{-4}b^{*}$ and assuming $g=1$, we determine the
contact overstress $\eta _{\rm cnt}$~= 1.2~meV. From our fit we find the cut off value $\eta_{t}$~= 0.24~meV  in a  good agreement with the value $\eta_{t}$~= 0.25~meV derived from the  persistent strain $q_{t}=1.10^{-4}$~\AA$^{-1}$ derived from pc data [1]. A comparison of the passive and
active conversion mechanisms is given by the ratio
$R_{p}/R_{a}j_c\approx 1.2$ determined from our fit, hence the two contributions are comparable.

\begin{figure}[thb]
\includegraphics{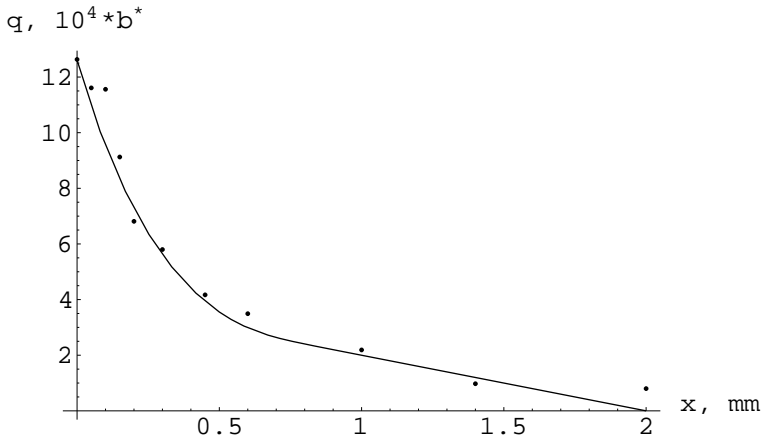}
\epsfxsize \hsize
\caption{ $2q(x)$ versus experimental points from [1].}
\end{figure}

In Fig.3 we  present calculations for the  configuration of [3]  within the same model and with the same parameters as for [1].
Now the distance is short $L=670\mu m$ and the current is high $j_{tot}=5j_{t}$. We cannot detect any visible difference between the
simulations with the cut off $\eta_t$ and without it. The pronounced bulk gradients observed for these short distances between contacts disappear completely when we increase this spacing  2.5 times which is just the rescaling from multi-contact to X-ray geometries. Hence in experiments [3]
 the whole length is affected by the contacts, the partial currents are far from equilibrium, conversion goes on over the whole sample but is far from being completed.
\begin{figure}[thb]
\includegraphics{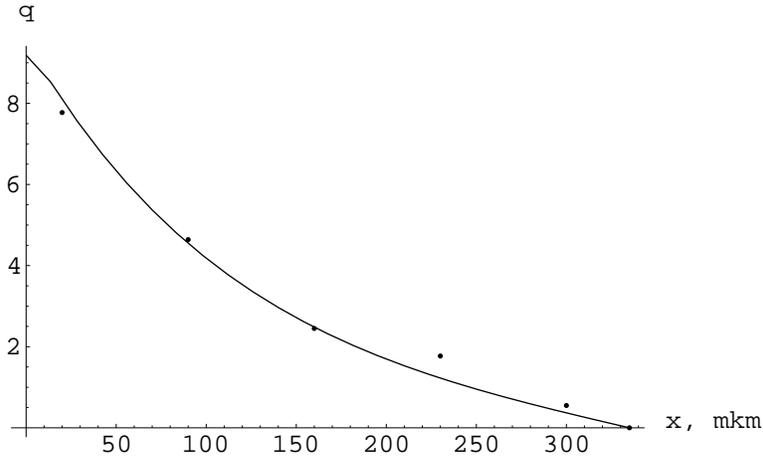}
\epsfxsize \hsize
\caption{
$q(x)$  versus  data from [3].}
\end{figure}

Clearly  in X-ray [1,2]  and in multi-contact  [3] experiments the two completely different regimes have been observed.
Both experiments [1,2]  recover the {\em{quenched gradients of the same origin, but not the phase slip stress from contacts}} as it has been supposed.

Our fitting allows to estimate a mean concentration (per sample cross-section)
of the lattice faults $N_{flt}$ which provide the active conversion. Indeed
$j_c^\prime=2{\cal P}N_{flt}j_c$\ where ${\cal P}$ is the probability to create a D-line at one fault over-passing each DW period. Our hypothesis 
$R=R_a N_F^i \eta j_c$ means that 
${\cal P}(\eta )=\eta /\eta ^{*}$ where the constant $\eta ^{*}$ is determined by
the condition ${\cal P}(\eta ^{*})\sim 1$. 
Hence $\eta ^{*}$ can be estimated as the highest energy payed to create a D-loop which corresponds to just one layer from the surface. 
From the characteristic energy of interplane interactions  we expect 
$\eta ^{*}\sim 100K$ so that
$N_{flt}\approx 10^{-5}A^{-1}$.
We have obtained the distance $L_{\rm flt}$~=
 $1/N_{\rm flt}\sim 10~\mu$m which is less but comparable to the initial decay
scale $(q/q^\prime)_{cnt} \approx 60~\mu$m and 30 times below the total
characteristic length $\approx 300~\mu$m. We see that the current conversion
is provided by about $10^1-10^2$ defects with a maximal (near the contact)
characteristic efficiency of few procents: $\eta_{cnt}/\eta^{*}\sim
0.1$. We conclude that D-loops of opposite signs
predominantly annihilate each other by exchanging locally the normal
currents and only  some fraction
 survives to proliferate across the sample.
\vskip .2in
{\bf 4. DISCUSSION AND CONCLUSIONS.}\\
\vskip .1in
In sliding state the collective current  which can reach its nominal value  provided that the equilibration between the normal  and the collective components is completed. In contradiction to earlier theoretical expectations and interpretations of experiments, we show that this state is mainly undeformed. 
Namely, elastic deformations - either from the
distributed stress [2,10]  or from the phase slip tension at the contacts [3] are incompatible with an equilibrium with respect to the conversion. 

We conclude (see also [1]) that for a long enough sample the large $q$-values near the
contacts fall nearly exponentially towards zero in accordance with the $HeteroN$ scenario. The detailed fit requires for the contribution of the active conversion generated by the sliding itself.  The large gradients
observed in earlier multi-contact studies [3] can now be
re-interpreted as due to the  size effect:  the distance between
contacts is so short that only a small part of the applied normal
current is converted, and the currents $j_c$ and $j_n$ stay far
from their equilibrium values.  In this limit the strain $q(x)$ is
close to linear with only small contact increments.  In contrast,
earlier x-ray experiments [2], performed on longer samples
(4.5~mm) could not yet resolve the contact region and what was observed as
a bulk gradient coincides with remnant distortion $q^\prime_\infty$
from pinned DLs as revealed by [1].

Till now the DW abilities to slide and to be deformed were supposed to be
related. Our results show that they rather exclude each other.

\vskip.09in
{\bf References}\\
\vskip .09in

\begin{tabular}{rl}
1. &H.Requardt et al, PRL,  {\bf 80}, 5631 (1998) and in this volume.\\
2. &D.DiCarlo et al., Phys. Rev. Lett. {\bf 70}, 845 (1993).\\
3. &T.L.~Adelman et al., Phys. Rev. B {\bf 53}, 1833 (1996).\\
4. &S.G. Lemay {\it et al.}, Phys. Rev. B {\bf 57},
12781 (1998).\\
5. & M.E.~Itkis, et al, Phys. Rev. B {\bf  52}, R11545 (1995).\\
6. &M.E.~Itkis et al., J.Phys.: Condens. Matter 
{\bf 5}, 4631 (1993).\\
7. &S. Brazovskii and S. Matveenko, 
{\em J. de Physique. I.}, {\bf 1},269 \& 1173 (1991); \\
 &ibid {\bf 2}, 409 \& 725 (1992).\\
8. &S.Brazovskii, J. Physique I {\bf 3}, 2417 (1993); 
also  in {\it Electronic Crystals},\\ 
 &S.Brazovskii, P.Monceau eds,
 J. Physique IV {\bf 3} C2 (1993).\\
9. &S. Artemenko, A. Volkov, 
in  {\it Charge Density Waves in Solids}, 
 L.Gor'kov, G.~Gr\"{u}ner eds,\\ 
 &Modern Physics in Condensed Matter Science Vol.~25 
(North Holland, 1989).
 p. 365.\\
10.  &S. Brazovskii in  {\em Proceedings of the   NATO Summer School, 
Les Houches 95}, \\
 &C. Schlenker, M. Greenblatt eds., World Sci. Publ., 1996.\\
11.  &S. Brazovski in {\it Charge Density Waves in Solids}, 
L.P.~Gor'kov and G.~Gr\"{u}ner eds, \\
 &Modern Physics in Condensed Matter Science Vol.~25.\\ 
12.  &S.Ramakrishna et al., 
Phys. Rev. Lett. {\bf 68}, 2066 (1992).\\
13. &K.Maki, Phys. Lett. A {\bf 202}, 313 (1995).\\
14. &S.Brazovskii, A.Larkin, Synth. Met.,  {\bf 86}  (1997) 2223 and in this volume. \\

\end{tabular}
\end{document}